\documentclass[final,5p,times,twocolumn]{elsarticle}

\usepackage{lipsum}

\usepackage{lineno}
\usepackage{amssymb}
\usepackage{amsmath}
\usepackage{here}
\usepackage{braket}
\usepackage{bm}
\usepackage[driverfallback=dvipdfm,colorlinks=true, bookmarks=true,bookmarksnumbered=true,bookmarkstype=toc]{hyperref}
\usepackage{braket}

\usepackage[dvipsnames]{xcolor}

\journal{Nuclear Instruments and Methods in Physics Research Section A: Accelerators, Spectrometers, Detectors and Associated Equipment}

\begin{document}

\begin{frontmatter}

%% Title, authors and addresses

%% use the tnoteref command within \title for footnotes;
%% use the tnotetext command for theassociated footnote;
%% use the fnref command within \author or \address for footnotes;
%% use the fntext command for theassociated footnote;
%% use the corref command within \author for corresponding author footnotes;
%% use the cortext command for theassociated footnote;
%% use the ead command for the email address,
%% and the form \ead[url] for the home page:
%% \title{Title\tnoteref{label1}}
%% \tnotetext[label1]{}
%% \author{Name\corref{cor1}\fnref{label2}}
%% \ead{email address}
%% \ead[url]{home page}
%% \fntext[label2]{}
%% \cortext[cor1]{}
%% \address{Address\fnref{label3}}
%% \fntext[label3]{}

\title{Two-dimensional beam profile monitor for the detection of alpha-emitting radioactive
isotope beam}
\author[tohoku]{K.S. Tanaka}\corref{Tanaka}
\cortext[Tanaka]{present address:  Paul Scherrer Institute (PSI), Switzerland.}
\ead{tanaka@kaduo.jp}
\author[tohoku]{U. Dammalapati\corref{umakanth}}
\cortext[umakanth]{present address:  Physics section, DESM, RIE Mysuru, Karnataka, India.}
\author[tokyotech]{K. Harada}
\author[riken]{T. Hayamizu}
\author[tohoku]{M. Itoh}
\author[tohoku]{H. Kawamura}
\author[cns]{H. Nagahama}
\author[cns]{K. Nakamura}
\author[utokyo]{N. Ozawa}
\author[cns]{Y. Sakemi}
 
\address[tohoku]{Cyclotron and Radioisotope Center(CYRIC), Tohoku University, 6-3 Aramaki-aza Aoba, Aoba-ku, Miyagi 980-8578, Japan}
\address[tokyotech]{Department of Physics, Tokyo Institute of Technology, 2-12-1 Ookayama, Meguro-ku, Tokyo 152-8551, Japan}
\address[riken]{Nishina Center for Accelerator-Based Science, RIKEN, 2-1 Hirosawa, Wako, Saitama 351-0198, Japan}
\address[cns]{Center for Nuclear Study, Graduate School of Science, The University of Tokyo, 2-1 Hirosawa, Wako, Saitama 351-0198, Japan}
\address[utokyo]{Graduate School of Science, The University of Tokyo, 7-3-1, Hongo, Bunkyo-ku, Tokyo 113-0033, Japan}

\begin{abstract}
%% Text of abstract
Ions with similar charge-to-mass ratios cannot be separated from existing beam profile monitors (BPMs) in nuclear facilities in which low-energy radioactive ions are produced due to nuclear fusion reactions. In this study, we developed a BPM using a microchannel plate and a charge-coupled device to differentiate the beam profiles of alpha-decaying radioactive isotopes from other ions (reaction products) produced in a nuclear reaction. This BPM was employed to optimize the low-energy radioactive francium ion ($\rm{Fr}^{+}$) beam developed at the Cyclotron and Radioisotope Center (CYRIC), Tohoku University, for electron permanent electric dipole moment (e-EDM) search experiments using Fr atoms. We demonstrated the performance of the BPM by separating the $\rm{Fr}^{+}$ beam from other reaction products produced during the nuclear fusion reaction of an oxygen ($^{18}\rm{O}$) beam and gold ($^{197}\rm{Au}$) target. However, as the mass of Au is close to that of Fr, separating the ions of these elements using a mass filter is a challenge, and a dominant number of $\rm{Au}^{+}$ renders the $\rm{Fr}^{+}$ beam profile invisible when using a typical BPM. Therefore, by employing the new BPM, we could successfully observe the $\rm{Fr}^{+}$ beam and other ion beams distinctly by measuring the alpha decay of Fr isotopes. This novel technique to monitor the alpha-emitting radioactive beam covers a broad range of lifetimes, for example, from approximately 1 s to 10 min, and can be implemented for other alpha-emitter beams utilized for medical applications.
\end{abstract}
\begin{keyword}
 Microchannel plate (MCP) \sep Beam profile monitor \sep Charge coupled device (CCD) \sep Fr atom %keywords here, in the form: keyword \sep keyword

%% PACS codes here, in the form: \PACS code \sep code

%% MSC codes here, in the form: \MSC code \sep code
%% or \MSC[2008] code \sep code (2000 is the default)

\end{keyword}

\end{frontmatter}

%% main text
\section{Introduction}
An experiment to measure the electron permanent electric dipole moment (e-EDM)~\cite{Andreev2018} by employing ultracold radioactive isotopes of francium (Fr) is being set up at the Cyclotron and Radioisotope Center (CYRIC), Tohoku University~\cite{Sakemi_2011}. Considering theoretical calculations, among the paramagnetic atoms, the Fr atom exhibits a large enhancement factor $\sim 900$~\cite{PhysRevA.88.042507,Shitara2021}, that is, the ratio of atomic EDM to e-EDM. Moreover, laser cooling and trapping have been demonstrated for Fr isotopes~\cite{SPROUSE1997370}.

\par
\begin{figure}[htbp]
  \begin{center}
  \includegraphics[width=\hsize]{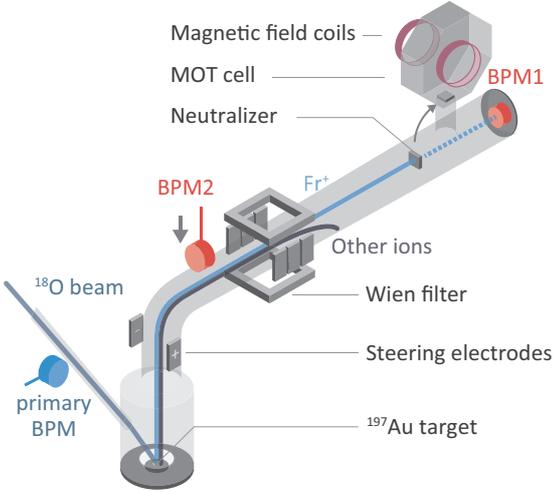}
  \caption{Schematic view of low-energy Fr ion ($\rm{Fr}^{+}$) beam line at CYRIC. $\rm{Fr}^{+}$ produced by surface ionization at the Au target passes through the Wien mass filter and is transported to the neutralizer yttrium target of dimensions 10~mm $\times$ 10~mm. Two beam profile monitors (BPMS) are used. BPM1 installed at the end of the beam line immediately after the neutralizer target, and BPM2 is placed before the Wien filter after bending the ion beam by $90^{\circ}$ with the steering electrodes (SE). The MOT cell is oriented $90^{\circ}$ (upward) to the beam line. The BPM for monitoring the $^{18}\rm{O}$ beam is also shown.}
  \label{CYRIC-beamline.eps}
  \end{center}
  \end{figure}

Schematics of the apparatus developed at CYRIC for Fr production and the low-energy beam line for the transportation of $\rm{Fr}^{+}$ are shown in Figure~\ref{CYRIC-beamline.eps}. $^{209-215}\rm{Fr}$ isotopes are produced by the nuclear fusion evaporation reaction between the oxygen ($^{18}\rm{O}$) beam from the AVF cyclotron and gold ($^{197}\rm{Au}$) target~\cite{published_papers/21027008}. The production rate of Fr is approximately $10^{6} /\mathrm{s}$, corresponding to a $^{18}\rm{O}$ beam current of $\approx$ 1~$\mu$A. The details of Fr isotope production and transportation are given in~\cite{Kawamura2015}. The produced Fr isotopes are ionized by surface ionization on the Au target and transported approximately 12 m away from the production region to the measurement area, in the absence of background radiation such as neutrons and gamma rays that are usually produced during nuclear reactions. The extracted $\rm{Fr}^{+}$ beam is focused, stopped, and accumulated on a neutralizer yttrium target depending on the lifetime of the Fr isotopes. Upon heating the yttrium foil to approximately $10^{3}$~K~\cite{Kawamura2015}, thermal Fr atoms are released, which are then collected in a magneto-optical trap (MOT)~\cite{Harada_2016}. These laser-cooled Fr atoms will be employed for the measurement of e-EDM in an optical dipole trap~\cite{PhysRevA.93.043407,JPSConf.Proc.2.010112}. 

\begin{figure}[htbp]
  \begin{center}
    \includegraphics[width=0.6\hsize]{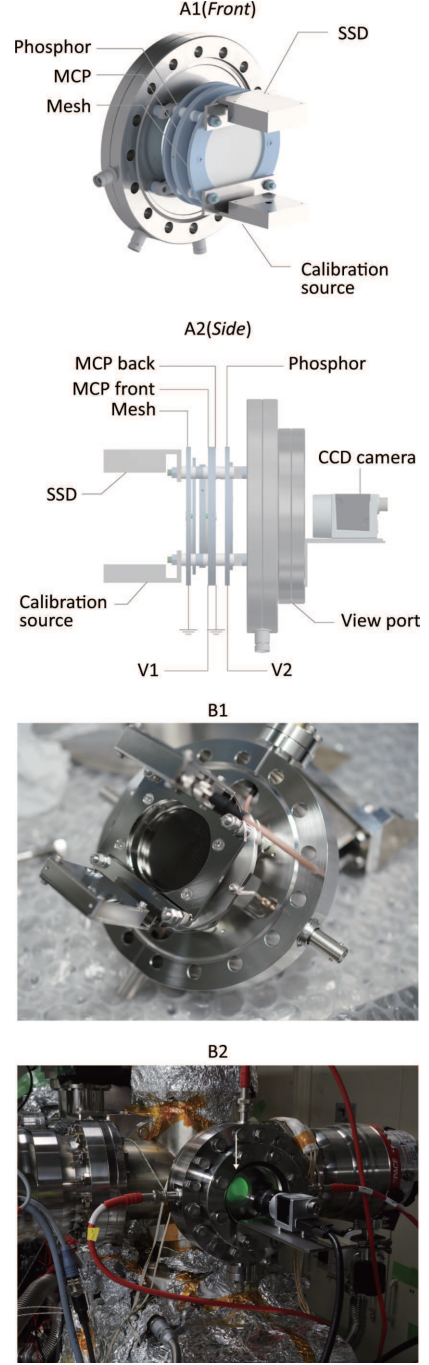}
    \caption{ {\bf A1-A2)}:Computer-aided design (CAD) diagram of BPM1. The BPM is assembled on a view port and consists of two sections: A1(\textit{Front}) including the  mesh, the MCP, and the phosphor screen stacked parallel to each other, and A2(\textit{Side}) including a charge-coupled device camera that is used to observe the fluorescence on the phosphor screen. The silicon-based solid-state detector (SSD) and radioactive source ($^{241}\rm{Am}$) are installed in front of the BPM to detect the alpha particles emitted from the surface of the MCP. The SSD provides the number of alpha decays from each isotope depending on the energy of the alpha particles. The voltage difference between the MCP front($V_{1}$) and its back (ground voltage) defines the gain (1000 $\sim$ 2000~V), while the voltage difference between the MCP back (ground voltage) and phosphor screen($V_{2} = 2000$~V) contributes to the focusing of the electron beam on the phosphor screen and determines the sharpness of the beam profiles. {\bf B1)}: Photograph of BPM1 before installation in the beam line. {\bf B2)}: Irradiation of BPM1 by rubidium ion beam. The green glow of the phosphor screen (indicated by the white arrow) can be observed when the Rb$^{+}$ beam impinges on it.}
    \label{bpm-cad-photo2.eps}
  \end{center}
\end{figure}

\par
In the Fr EDM experiment, the sensitivity of the EDM measurement depends on the number of accumulated atoms \cite{Sakemi_2011}.
In addition to the production and extraction efficiencies of Fr isotopes from the Au target, the efficient transportation of the low-energy $\rm{Fr}^{+}$ beam and separation of the Fr isotopes from other reaction products are important.
During the experiment, $\rm{Fr}^{+}$ is transported along with several other ions, including $\rm{Au}^{+}$ from the Au target. The yield of the $\rm{Au}^{+}$ beam is approximately $10^{9} /\mathrm{s}$, which is 1000 times higher than that of $\rm{Fr}^{+}$ ($10^{6} /\mathrm{s}$). Moreover, separating $\rm{Au}^{+}$ with a Wien mass filter is difficult because it has a charge-mass ratio similar to that of Fr. In addition, the Fr beam profile cannot be observed, because it is hidden in the other dominant ion beam profiles~\cite{published_papers/21027008}. Thus, there exists no (proper) diagnosis system to measure the beam profiles in the low-energy beam line, which is one of the major bottlenecks in the optimization of the secondary beam. 

\par
In this study, BPMs were developed to improve the beam transportation efficiency of a low-energy radioactive ion beam, that is, the ratio of the number of Fr ions reaching the neutralizer yttrium target to the number of Fr production and beam purity (ratio of $\rm{Fr}^{+}$ and other ions) of the radioactive ion beam between the Au target and the neutralizer yttrium foil, by measuring the beam profiles. To realize this BPM, a 2D BPM was developed using a microchannel plate with a phosphor screen and a charge-coupled device (CCD) camera for the alpha emitting $\rm{Fr}^{+}$ beam, to separate it from other reaction products. In the following sections, the design, construction, and characterization of the MCP-based BPM (Section 2), performance evaluation of the BPM (Section 3), and its application to measure the Fr beam profile (Section 4) are described. 

\section{Design/Construction of the beam profile monitor}
Microchannel plates (MCPs) are versatile devices that are used for various applications in diverse fields. The working principle, details, and applications of MCPs are described in~\cite{FRASER200213}. They have been used as beam position monitors for low-energy ($\sim$ keV) ion beams~\cite{doi:10.1063/1.1150059, LIENARD2005375}. 
Kinetic emission is the dominant effect observed when using MCPS, and the electron yield produced in this process depends on the velocity rather than on the ion kinetic energy~\cite{LIENARD2005375}.
In addition, a position calibration method for MCPs was also demonstrated to detect alpha particles and ions with energies of a few kiloelectronvolt (keV)~\cite{HONG201642}. Here, MCPs were implemented to detect ion beams and alpha particles from the decay of Fr isotopes, from which information of the beam profiles of the Fr$^{+}$ beams are obtained.

Two BPMs (BPM1 and BPM2) were constructed and installed in the low-energy beam line, each consisting of a microchannel plate (MCP), phosphor screen, and CCD camera. CAD drawings and pictures of the assembled BPMs are shown in Fig.~\ref{bpm-cad-photo2.eps}(A1-B2). The MCP (PHOTONIS APD-2-MA 40/32/25/8 D 60:1 NR, TC) is a chevron (two plates mounted) that has a sensitive region with a diameter 40~mm and pore diameter of 10~$\mu$m. The phosphor screen is an RHEED screen (ARIOS SG63-2). The impact of the ion beams on the MCP produced a cascade of electrons that propagate through one of the small channels by applying a strong electric voltage (1--2~kV) across the MCP. 
The electron clouds were converted into visible light by the phosphor screen, and the emitted light is observed using a CCD camera (Basler acA2500-14um, resolution: 2592 $\times$ 1944, fps: 14~Hz) for BPM1 and another camera (Basler acA1300-60gm, resolution: 1280 $\times$ 1024, fps: 60~Hz) for BPM2.

BPM1 was placed at the end of the low-energy beam line behind the neutralizer yttrium target (see Fig.~\ref{CYRIC-beamline.eps}). The profile of the ion beam at BPM1 could be observed on the CCD camera positioned at the rear of BPM1. During the beam profile measurement, the neutralizer was rotated 90$\rm^{o}$ with respect to the ion beam path, and the magnetic field coils of the MOT cell were switched off. Therefore, the profile of the beam from the neutralizer could be geometrically reconstructed. Further, the BPM2 was positioned before the Wien mass filter to optimize the Fr production rate on the Au target. The light emitted from the phosphor screen of BPM2 incidents on a mirror installed at 45$\rm^{\circ}$ with respect to the ion beam path and was observed through a view port. 

Two types of measurements were performed to optimize the performance of the BPMs. The first uses a rubidium ion (Rb$^{+}$) beam, and the other employs the Fr$^{+}$ beam. The Rb$^{+}$ beam measurements are primarily used to determine the voltage characteristics of the MCPs and to optimize their performance. The results are presented in Section~\ref{section-performance-evaluation-of-BPMs}. To optimize the performance of BPM for Fr, initial measurements on BPM1 were performed using a silicon-based solid-state detector (SSD) and a radioactive source ($^{241}\rm{Am}$). They were installed in the beam line for calibration as well as for the detection of alpha particles emitted from the surface of the MCP. The SSD enables the counting of the number of alpha decays from each isotope depending on the energy of the alpha particles. Furthermore, the solid angle for the detection of alpha particles via SSD was 0.25 sr, corresponding to a detection efficiency of 0.2\%.

When alpha-emitter ions such as Fr are injected into the BPM, these ions accumulate on the surface of the MCP and decay, emitting alpha particles. Because the energy of the alpha particles is higher (6.5~MeV from $^{210}\rm{Fr}$) compared to the kinetic energy of the ion beams (1~keV), the number of cascade electrons produced is comparatively large. Therefore, the observation of the small flux of the Fr$^{+}$ beam ($<$~100 ions/s ) in the background of high flux of other ions is easier. In addition, if the ion beam is stopped for a sufficient time, that is, the time corresponding to the lifetime of Fr isotopes, the profile of the Fr beam can only be observed, which is due to the accumulated Fr isotopes on the surface of the MCP.

\section{Performance evaluation of BPMs}\label{section-performance-evaluation-of-BPMs}

\subsection{Dependence of MCP voltage on luminosity from BPMs: Rb$^{+}$ beam}
\begin{figure}[htbp]
  \begin{center}
    \includegraphics[width=\hsize]{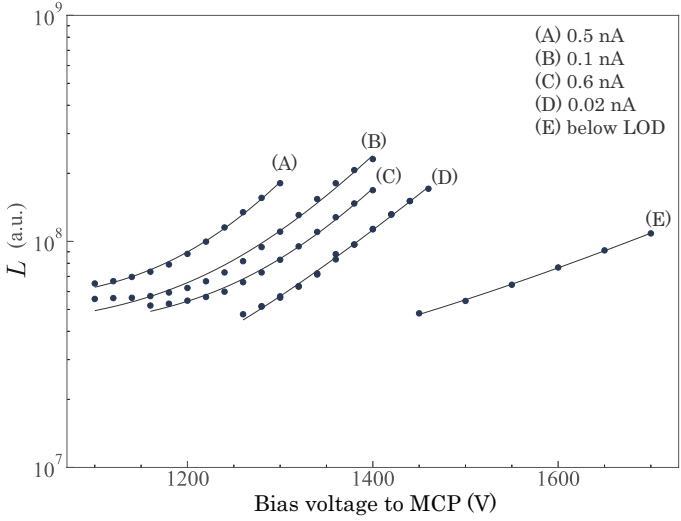}
    \caption{The variation of parameter $L$ as a function of the MCP front voltage is measured using a rubidium ion beam for five different beam fluxes (A)--(E). The measurement details are provided in the text. The MCP front voltage was adjusted to not saturate the CCD under different flux conditions. The ion beam flux was measured at the neutralizer position with a Faraday cup and a picoammeter (Keithley 6485/J). LOD --- Limit of Detection.}
    \label{mcp-front-dependence.eps}
  \end{center}
\end{figure}

Voltages were applied to three plates in the MCP: MCP front, MCP back, and phosphor screen (see Fig.~\ref{bpm-cad-photo2.eps}). The voltage difference between the MCP front and its back defines the gain, while the voltage difference between the MCP back and phosphor screen contributes to the focusing of the electron beam on the phosphor screen and determines the sharpness of the beam profiles. 
However, the voltage difference between the MCP front and back should be adjusted such that saturation does not occur on the CCD camera, and the maximum operating voltage is 2000~V.
In addition, the voltage difference between the MCP back and the phosphor screen was 2000~V. 
Because the gain of the MCP has an exponential relation with the applied MCP front voltage, the sum of luminous intensity at the region of interest (ROI) on the CCD camera ($L$) is expressed as 
\begin{equation}
  L = A \exp{(kV_{1})}+B,
  \label{eq:relationship-intensity-votalge}
\end{equation}
where $V_{1}$ is the voltage between MCP front and back, and $k$, $A$, and $B$ are constants.
Figure~\ref{mcp-front-dependence.eps}(A)--(E) shows the relationship between the parameter $L$ and $V_{1}$ measured with the Rb ion beam at five different beam fluxes. 
The mesh and MCP back are set to ground voltage, the MCP front voltage was varied from 1000~V to 2000~V. At the same time care was taken to avoid saturation of the CCD for different beam flux conditions.

The parameter $k$ was obtained after fitting each line with Eq.\ref{eq:relationship-intensity-votalge}. The respective values of $k$ are: (A) $k = 0.011$, (B) $k = 0.010$, (C) $k = 0.012$, (D) $k = 0.008$, (E) $k = 0.004$.
As an example, in Figure~\ref{mcp-front-dependence.eps}(A), measurements correspond to $V_{1}$ = 1100--1300~V, and the exponential relations were confirmed in this range. The MCP front voltage was varied in steps of 20~V covering a range of 200~V. Similarly, data was taken for voltage ranges shown in Figure~\ref{mcp-front-dependence.eps}(B)--(E). These measurements were performed at 5 different beam fluxes.

\subsection{Fr beam profile measurement}

The measurement sequence of the $\rm{Fr}^{+}$ beam profiles employing BPMs is shown in Fig.~\ref{bpm-sample.eps}:

  \begin{figure*}[htbp]
    \begin{center}
      \includegraphics[width=\hsize]{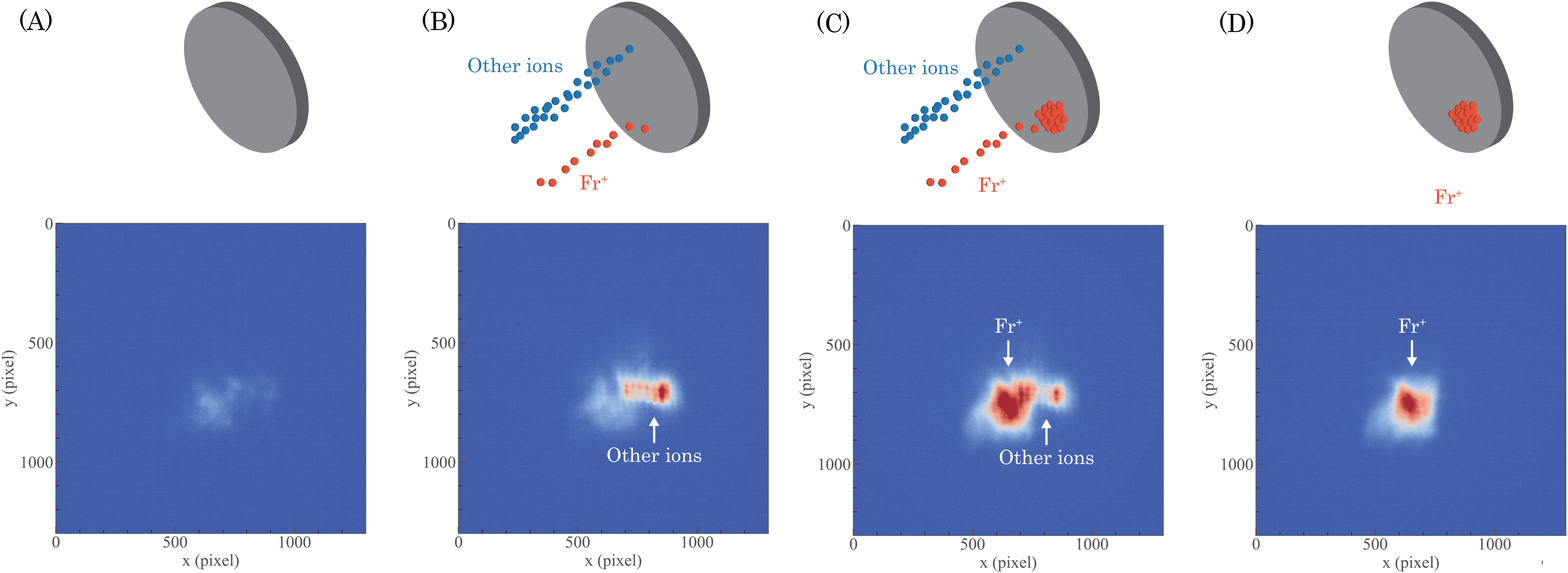}
      \caption{Typical measurement sequence for the observation of the Fr$^{+}$ beam profiles during the experiment. \textit{Top}: Schematic images of beam profile monitor (BPM) and ion beams (Red dots---$\rm{Fr}^{+}$ and Blue dots --- other ions). \textit{Bottom}: The measured images of $\rm{Fr}^{+}$ and other ions on BPM1 with MCP bias voltage: 1400~V. (A) The ion beam is stopped for sufficient time until the fluorescence of the alpha decay of Fr accumulated from the previous sequence disappears. (B) An ion beam is irradiated on the BPM. Initially, only the beam profile due to the impact of the other ion beams is observed because the dominant part of the beam profile is due to $\rm{Au}^{+}$ and other ions that pass through the mass filter in addition to $\rm{Fr}^{+}$. (C) After irradiation corresponding to the lifetimes of Fr isotopes (longer than 3 min), the beam profile due to the alpha decay of Fr is observed. (D) In step 4, beam irradiation is stopped, and only the Fr beam profile remains. In these measurements, the beam flux of $\rm{Fr}^{+}$ is approximately $2 \times 10^5$/s, and that of other ions is $4 \times 10^{7}$/s, respectively.}
      \label{bpm-sample.eps}
    \end{center}
  \end{figure*}

  \begin{figure}[H]
    \begin{center}
      \includegraphics[width=\hsize]{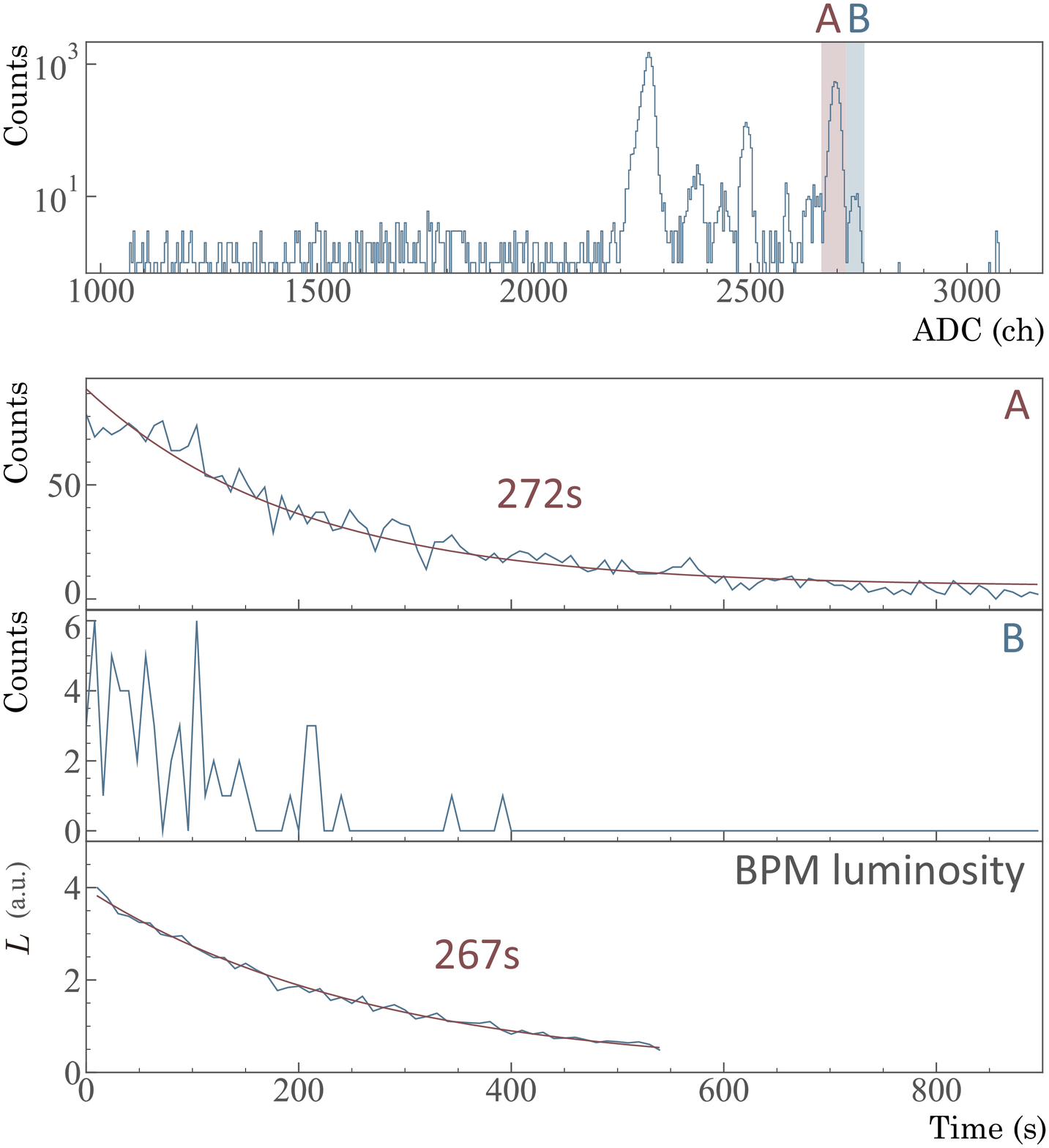}
      \caption{
  The energy spectrum of the alpha decay measured by the SSD (top) obtained by BPM1. Peak A (red) corresponds to the energy of alpha decay from $^{210}\rm{Fr}$ and $^{211}\rm{Fr}$. Peak B (blue) corresponds to that from $^{208}\rm{Fr}$ and $^{209}\rm{Fr}$. Other peaks are from daughter isotopes of Fr or $^{241}\rm{Am}$ --- a test source.
        A and B show the time structure of the SSD counts after stopping the ion beam [(D) in Fig. \ref{bpm-sample.eps}].
        The lifetimes of $^{210}\rm{Fr}$ and $^{211}\rm{Fr}$ were estimated as $\tau =272(13)$~s by fitting with an exponential function (red line). $^{208}\rm{Fr}$ and $^{209}\rm{Fr}$ lifetimes were not estimated because of the lack of statistics.
        The CCD luminosity shows the time structure of total luminosity on BPM1. From this figure, the lifetime of the alpha-decaying isotope is estimated as $\tau = 267(15)$~s by fitting (red line), which is consistent with the lifetime of the SSD. The dominant part of the alpha decay is from $^{210}\rm{Fr}$ and $^{211}\rm{Fr}$ isotopes.}
      \label{run24-bpm-ssd-decay.eps}
    \end{center}
  \end{figure}

\begin{enumerate}[(A)]
 \item The ion beam is stopped for sufficient time till the fluorescence from the alpha decay of Fr accumulated from the previous sequence disappears.
\item Start the ion beam (consisting $\rm{Au}^{+}$, $\rm{Fr}^{+}$ and other ions) irradiation on the BPM (see Fig.~\ref{bpm-sample.eps}(B)). Here, the dominant part of the beam profile is due to $\rm{Au}^{+}$ and other ions, which also pass through the mass filter.
  \item After waiting ($\sim10$ min) for enough time corresponding to the lifetime of Fr isotopes, the Fr beam profile is observed, which, otherwise, is not observable with an impact on the MCP owing to the small number of Fr ions. However, the beam profiles are easily observed because of the alpha decay of Fr because they produce a large number of electrons in the MCP. Therefore, the Fr beam profile is gradually visible owing to its accumulation on the surface of the MCP.
  \item The ion beam irradiation is stopped (here, after 250~s). Immediately after the beam is extinguished, the beam profile of the other ion beams is observed, and they disappear quickly. Consequently, the remaining profile of the beam is only due to Fr$^{+}$.
\end{enumerate}
From the above-mentioned measurements, the lifetimes of the Fr isotopes were also estimated in addition to their beam profiles.

\subsection{Lifetimes of Fr isotopes}
The number of signal counts on the SSD is used for the absolute calibration of the sum intensity of the BPM. Figure \ref{run24-bpm-ssd-decay.eps} shows the SSD counts corresponding to the number of alpha decays measured at the surface of the BPM1 and the corresponding intensity on the CCD camera $L$ during the Fr beam experiment.
The accumulation of Fr on the surface of the MCP during beam irradiation, as well as the alpha decay of Fr isotopes after stopping the beam are observed using both the SSD and CCD camera.
The number, $N$ of $\rm{Fr}^{+}$ on the MCP is expressed as
\begin{align}
  \frac{\rm{d}N}{\rm{d}t} & = & f - \frac{1}{\tau}N,&     & (t_{0} < t <t_{1}, \mbox{ beam on)} \\
  \frac{\rm{d}N}{\rm{d}t} & = &  - \frac{1}{\tau}N,&     & (t_{1} < t, \mbox{ beam off)} 
\end{align}
where $t_{0}$ is the start time of the beam irradiation and $t_{1}$ is the stop time of the beam;
 $f$ is the Fr beam flux, and $\tau$ is the lifetime of the Fr isotope.
The number of Fr particles during beam irradiation ($N_{\rm{on}}$) and after stopping the beam ($N_{\rm{off}}$) are expressed as
\begin{align}
  N_{\rm{on}} & = & \tau f \left(1-\exp{\frac{t-t_{0}}{\tau}}\right) +N(t_{0}) \exp{(-\frac{t-t_{0}}{\tau})},  \label{eq:N_on}  \\
  N_{\rm{off}} & = & \left[ \tau f\exp{(\frac{t_{1}}{\tau})} + N(t_{0}) \exp{(\frac{t_{0}}{\tau})} \right] \exp{(-\frac{t}{\tau})}.\label{eq:N_off} 
\end{align}

The Fr beam flux, $f$, and the lifetime of Fr, $\tau$, can be estimated by fitting to Eq.~\ref{eq:N_on} and Eq.~\ref{eq:N_off}, respectively. 
Further, the data obtained from one of the measurements from the sequence described above are presented in Fig.~\ref{run24-bpm-ssd-decay.eps}. The lifetime of Fr was estimated by fitting Eq.~\ref{eq:N_off} to the measured data. Consequently, considering the intensity on the CCD camera, the estimated Fr lifetime was $\tau = 267(15)$~s and the sum lifetime of $^{210}\rm{Fr}$ and $^{211}\rm{Fr}$ from the SSD counts was $\tau = 272(13)$~s, respectively.

The Fr isotopes are produced by the fusion evaporation reaction $^{18}\rm{O}$ + $^{197}\rm{Au}$ $\rightarrow~^{215-x}\rm{Fr} + xn$, where $x$ is the number of evaporated neutrons.
During the reaction, several alpha-decaying Fr isotopes were produced. Of these, few are short-lived, such as, $^{208}\rm{Fr}$ ($\tau = 85.3$~s) and $^{209}\rm{Fr}$ ($\tau = 72.9$~s), while others are long-lived, such as, $^{210}\rm{Fr}$ ($\tau = 275.3$~s) and $^{211}\rm{Fr}$ ($\tau = 268.3$~s) \cite{CHIARA2010141}. In addition, long-lived daughter isotopes of Fr~\cite{STANCARI2006390} can be detected from both SSD and BPMs. Thus, the estimated lifetime from this measurement is the mixed lifetime of these isotopes.
In the case of SSD counts, the measured energy range of alpha particles is narrowed down to focus on the long-lived Fr isotopes. This partly explains why the lifetime of Fr from SSD counts is consistent with the mixed lifetime of $^{210}\rm{Fr}$ and $^{211}\rm{Fr}$~\cite{STANCARI2006390}. The beam profile of the longer-lived isotope can be neglected for Fr beam profile observation by background subtraction.

%\begin{figure}[htbp]
%  \begin{center}
%    \includegraphics[width=\hsize]{./figures/relation-ssd-bpm.eps}
%    \vspace*{8mm}
%    \caption{A correlation plot between the SSD counts and intensities of CCD camera in Fig.~\ref{run24-bpm-ssd-decay.eps.}}{\bf(Need to be checked again)}
%    \label{relation-ssd-bpm.eps}
%  \end{center}
%\end{figure}
\begin{figure*}[tb]
  \begin{center}
    \includegraphics[width=\hsize]{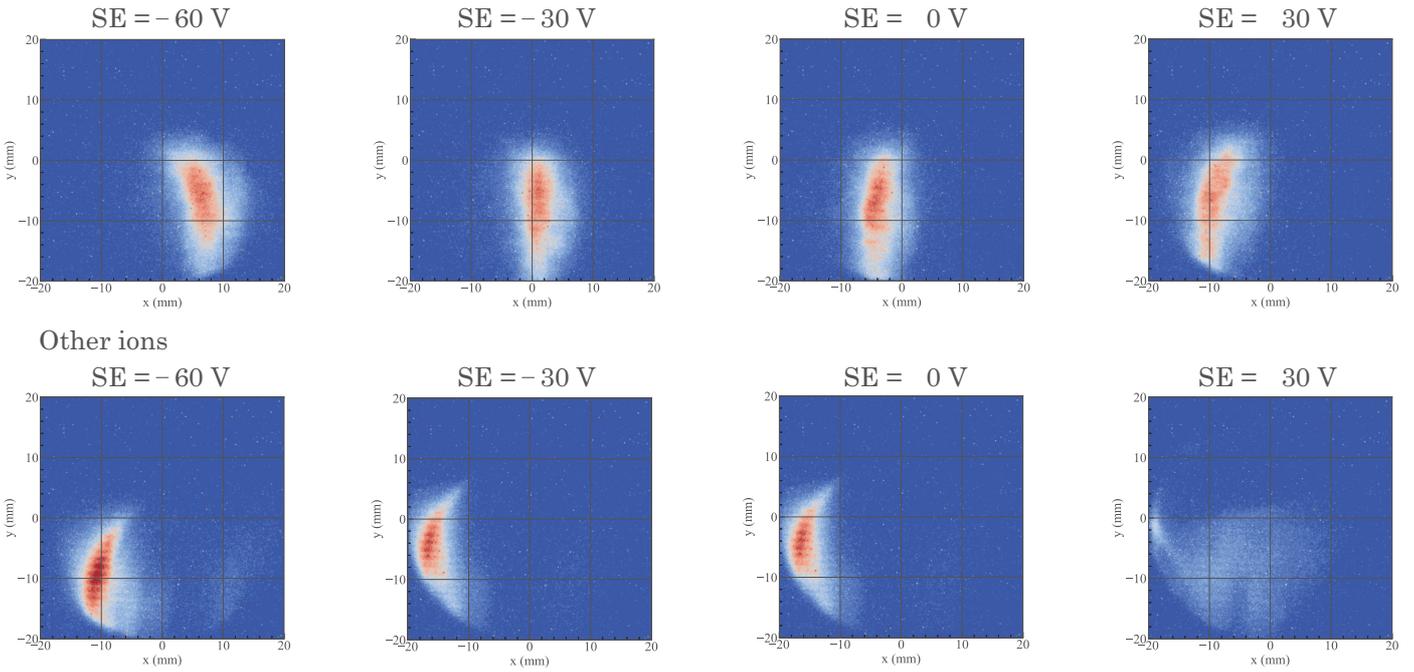}
    \caption{Beam profiles of alpha decaying Fr isotopes and other ions measured at BPM2 with SE=$-60$~V, SE=$-30$~V, SE=$0$~V, SE=$30$~V applied to the steering electrodes located after the Fr production region.}
    \label{bpm-st-dependence.eps}
  \end{center}
\end{figure*}

\begin{figure}[htbp]
  \begin{center}
    \includegraphics[width=\hsize]{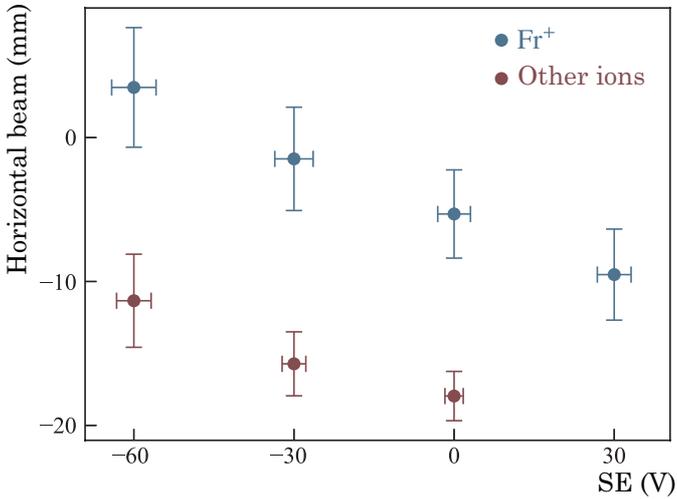}
    \caption{Variation of the horizontal position of $\rm{Fr}^{+}$ (blue) and other ion beams (red) as a function of steering electrodes voltage (SE). The beam positions of both Fr and other ions varied as a function of the voltage applied to the steering electrodes. For 30~V applied to steering electrodes, the Fr beam position could not be identified because the beam profile was beyond the detection range of MCP.}
    \label{st-dependence-plot.eps}
  \end{center}
\end{figure}

\begin{figure}[htbp]
  \begin{center}
    \includegraphics[width=\hsize]{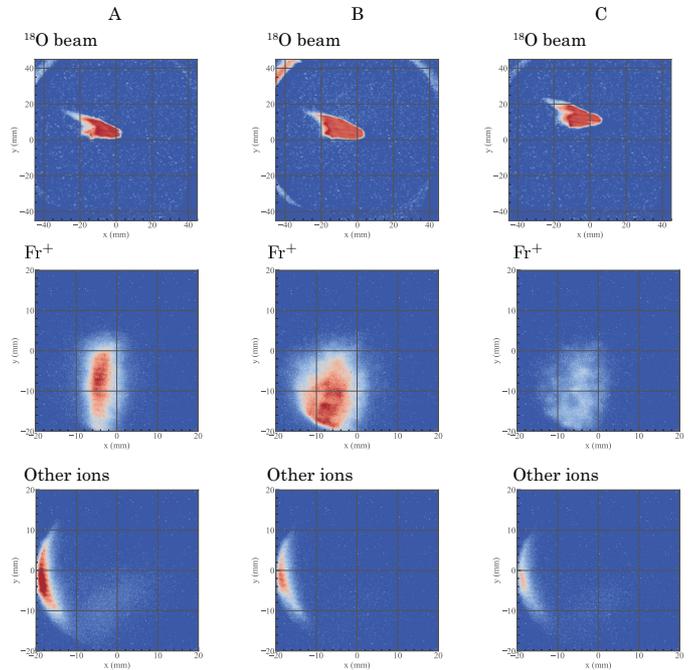}
    \caption{Beam profiles of $\rm{Fr}^{+}$ and other ions with three different conditions of primary $^{18}\rm{O}$ beam (A -- C). Depending on the primary beam position, the beam position of Fr varied and moved down. The beam positions of the other ions are not covered by the sensitive area of the profile monitor.}
    \label{primary-beam-dependence-plot.eps}
  \end{center}
\end{figure}

\section{Application of the BPM for beam transport}

\subsection{Beam control test with steering electrodes}
To confirm the beam diagnosis obtained using BPMs, the ion beam profiles at BPM2 were measured by applying different voltages to the steering electrodes (SE) (see Fig.~\ref{CYRIC-beamline.eps}).
In Fig.~\ref{bpm-st-dependence.eps} measured beam profiles of the ions at BPM2 are shown for SE=$-60$~V, SE=$-30$~V, SE=$0$~V, and SE=$30$~V, respectively, applied to the steering electrodes.
The variation in the horizontal beam position of $\rm{Fr}^{+}$ and other ions along with the voltages applied to the steering electrodes is shown in Fig.~\ref{st-dependence-plot.eps}.
The differences in the beam profiles between the alpha-particle-emitting isotopes and other ions originate from the initial beam size and shape in the reaction region on the Au target surface. It should be noted that Fr is produced at the collision point of the $^{18}\rm{O}$ beam and the Au target, whereas the other ions are evaporated from the Au target.

\subsection{Dependence of beam profile on $^{18}\rm{O}$ beam}
The beam profiles of both $\rm{Fr}^{+}$ and other ions depend on the position of the $^{18}\rm{O}$ beam colliding with the Au target which has a diameter of 16 mm. To verify this, a BPM (see Fig.~\ref{CYRIC-beamline.eps}) was placed immediately before the Au target to obtain the beam profile of the $^{18}\rm{O}$ beam. In Fig.~\ref{primary-beam-dependence-plot.eps}(A–C) beam profiles of $\rm{Fr}^{+}$ and other ions are shown for three different positions of the $^{18}\rm{O}$ beam. According to the beam position of $^{18}\rm{O}$, the beam position of $\rm{Fr}^{+}$ varied and moved down (A--C). However, the beam positions of the other ions are not covered by the sensitive area of the BPM because of the optimization of the purity of the $\rm{Fr}^{+}$ beam.

\section{Conclusion}
In conclusion, a BPM based on an MCP with a phosphor screen and a CCD camera was developed for the detection and separation of alpha-emitting radioactive isotope beams from other reaction products produced in a nuclear reaction. The BPM was characterized using a radioactive source and a rubidium beam. To the best of our knowledge, a BPM has been successfully implemented for the observation of the $\rm{Fr}^{+}$ beam profile, by observing the alpha decay of Fr isotopes for the first time. In addition, the secondary ion beam optimization was achieved by removing other ions from the $\rm{Fr}^{+}$ beam, which is one of the critical aspects in the low-energy Fr beam line at CYRIC.
We are also developing a high intensity laser-cooled Fr factory at the RIKEN accelerator facility and the BPM is an important device to optimize the $\rm{Fr}^{+}$ beam \cite{doi:10.1063/5.0037134}.
This novel technique to monitor the alpha particle-emitting radioactive beams covered a range of lifetimes from 1 s to approximately 10 min. However, this is limited by the exposure time of the CCD after the ion beam is stopped. This method can be harnessed for medical applications where alpha-emitting radioactive beams are used in addition to AMO and other fields~\cite{Poty878}.

\section{Acknowledgment}
The experiment was performed under a user program proposal at CYRIC, and the authors wish to acknowledge the expertise and help of the technical staff of the CYRIC accelerator facility. This research work was supported by JSPS KAKENHI, [Grant Numbers 26220705 and 19H05601].
%% The Appendices part is started with the command \appendix;
%% appendix sections are then done as normal sections
%% \appendix

%% \section{}
%% \label{}

%% If you have bibdatabase file and want bibtex to generate the
%% bibitems, please use
%%
%%  \bibliographystyle{elsarticle-harv}
%%  \bibliography{<your bibdatabase>}

%% else use the following coding to input the bibitems directly in the
%% TeX file.

%%\begin{thebibliography}{00}
%%
%% \bibitem[Author(year)]{label}
%% Text of bibliographic item

%%\bibitem[ ()]{}

\bibliographystyle{unsrt}
  \bibliography{NIM-FrEDM-ProfileMonitor}

\end{document}